\documentclass[aps,prl,twocolumn,showpacs]{revtex4}

\usepackage{mathptm}    
\usepackage{dcolumn}                    
\usepackage{bm}                         
\usepackage{graphicx}

\newcommand {\ba} {\begin{eqnarray}}
\newcommand {\ea} {\end{eqnarray}}

\newcommand {\be} {\begin{equation}}
\newcommand {\ee} {\end{equation}}

\usepackage{times}

\begin{document}

\title{Volovik effect in the $\pm$s-wave state for the iron-based superconductors}

\author{Yunkyu Bang}
\email[]{ykbang@chonnam.ac.kr} \affiliation{Department of Physics,
Chonnam National University, Kwangju 500-757, Republic of Korea\\
and Asia Pacific Center for Theoretical Physics, Pohang 790-784,
Republic of Korea}

\begin{abstract}
We studied the field dependencies of specific heat coefficient
$\gamma(H) = \lim_{T \rightarrow 0} C(T,H)/T $ and thermal
conductivity coefficient $\lim_{T \rightarrow 0} \kappa(T ,H)/T$
of the $\pm$s-wave state in the mixed state. We found that  it is
a generic feature of the two band s-wave state with the unequal
sizes of gaps, small $\Delta_S$ and large $\Delta_L$, that Doppler
shift of the quasiparticle excitations (Volovik effect) creates a
finite density of states, on the extended states outside of vortex
cores, proportional to $H$ in contrast to the $\sqrt{H}$
dependence of the d-wave state. Impurity scattering effect on the
$\pm$s-wave state, however, makes this generic $H$-linear
dependence sublinear approaching to the $\sqrt{H}$ behavior.
Our calculations of $\lim_{T \rightarrow 0} \kappa(T ,H)/T$
successfully fit the experimental data of Ba(Fe$_{1-x}$Co$_x )_2$
As$_2$ with different Co-doping $x$ by systematically varying the
gap size ratio $R= |\Delta_S | / |\Delta_L |$. We also resolve the
dilemma of a substantial value of $\gamma(H \rightarrow 0)$ but
almost zero value of $\lim_{T \rightarrow 0} \kappa(T ,H
\rightarrow 0)/T$, observed in experiments.
\end{abstract}


\date{\today}
\maketitle

{\it Introduction:} As to the pairing symmetry of the recently
discovered Fe-pnictide superconductors (SC) \cite{Kamihara}, the
$\pm$s-wave state \cite{Mazin,Bang-model} is considered as the
most natural and promising pairing state. Genuinely being an
s-wave gap state, this model is expected to provide exponentially
activated behaviors of various thermodynamic and transport
properties due to full gap(s) around Fermi surface(s) (FS) in the
superconducting phase.
Angle resolved photoemission spectroscopy (ARPES) experiments
\cite{photoemission-s} and early measurements of penetration depth
of $M$-1111 ($M$=Pr, Nd, Sm) \cite{1111} and (Ba,K)Fe$_2$As$_2$
\cite{Ba-122} confirmed this expectation. However, some power laws
of temperature dependence measured, for example, with the nuclear
spin-lattice relaxation rate $T_1 ^{-1} \sim T^3$ \cite{T1} as
well as the penetration depth $\lambda(T) \sim T^{2-2.5}$ on
Ba(Fe,Co)$_2$As$_2$ \cite{BaCo-122} were not consistent with this
simple picture. But these power law dependencies were successfully
explained theoretically with the $\pm$s-wave pairing state by the
sign-changing feature of the order parameter (OP) and the related
noble impurity effect \cite{Bang-imp,theory-T1} and hence
strengthened the status of the $\pm$s-wave state as the pairing
symmetry of Fe-pnictide superconductors.

However, recent measurements of specific heat coefficient
$\gamma(H)=C(H)/T$ and thermal conductivity coefficient
$\kappa(H)/T$ in the mixed state and their field dependencies are
posing a new challenge to the $\pm$s-wave pairing model. Several
measurements of specific heat coefficient showed a strong field
dependence: Ba$_{0.6}$K$_{0.4}$Fe$_2$As$_2$ \cite{SH_KBa122}
($\propto H$), (Fe$_{0.92}$Co$_{0.08}$)$_2$As$_2$
\cite{SH_CoBa122} (sublinear in $H$), and
LaO$_{0.9}$F$_{0.1-\delta}$FeAs \cite{SH_La1111} ($\propto
\sqrt{H}$).
And thermal conductivity measured on Co-doped Ba-122
\cite{kappa_CoBa122} and K-doped Ba-122\cite{kappa_KBa122} also
showed various field dependencies. In particular, Tanatar et al.
\cite{kappa_CoBa122} measured $\kappa (H)/T$ of Ba(Fe$_{1-x}$Co$_x
)_2$As$_2$ for several values of $x$ and found that the field
dependence of $\kappa (H)/T$ continuously evolves from
exponentially flat, near linear in $H$, and to near $\sqrt{H}$
behavior with increasing Co doping $x$.

Another puzzling observation is that, while the zero field limit
of $\gamma(H=0) / \gamma_{normal}$ in several experiments
\cite{SH_KBa122, SH_CoBa122, SH_La1111} showed a substantial
values, the zero field limit of $\kappa (H=0)/T$
\cite{kappa_CoBa122, kappa_KBa122} of the same compounds with
similar dopings approaches to negligibly small value. It implies
that there exist a large amount of zero energy excitations but
they do not contribute to thermal conductivity. Neither a nodal
gap state nor a simple isotropic s-wave gap state can be easily
compatible with these observations.

In this paper, we study the field dependence of the specific heat
and thermal conductivity coefficients of the $\pm$s-wave model in
the mixed state using a semiclassical approximation which causes
Doppler shift of quasiparticle excitations outside of vortex cores
(so-called "Volovik effect" \cite{Volovik}). Recently, Mishra et
al. \cite{Mishra} studied the field dependence of thermal
conductivity on the isotropic $\pm$s-wave model with equal size
gaps using a different method \cite{BPT} and concluded that this
model is incompatible with the experiments \cite{kappa_CoBa122,
kappa_KBa122}. Their result agrees with ours in that particular
case. In the current paper, however, we show that it is essential
to take into account the unequal sizes of gaps in the $\pm$s-wave
model. This feature is not only more realistic but also introduces
genuinely new physics to explain experimental data. We found that
the unequal size of two isotropic s-wave gaps causes the field
dependencies of $\kappa(H)/T$ as well as $\gamma(H)$ to be linear
in $H$ at the low field regime.

This generic $H$-linear dependence in the $\pm$s-wave model is in
contrast to the $\sqrt{H}$ dependence in the d-wave state
\cite{Volovik} although both are due to Doppler shift effect on
the extended states outside of vortex cores -- it should also be
contrasted to the $H$-linear contribution to specific heat from
the core bound states in a single s-wave state \cite{Hussey} that
cannot contribute to thermal conductivity. We then show that the
impurity scattering, combined with Volovik effect, make the
generic $H$-linear dependence gradually modified to become
sublinear and to approach to the $\sqrt{H}$ dependence.

{\it Formalism:} For the generic two band model of the $\pm$s-wave
pairing state, we assume two isotropic s-wave order parameters
(OPs) $\Delta_S$ and $\Delta_L$ of opposite signs on the two
representative bands of the Fe pnictide materials. $\Delta_S$ and
$\Delta_L$ represent a smaller and a larger size gap on the
corresponding bands, respectively. All energy units in this paper
is normalized by $\Delta_L$.

We use the semiclassical approximation where the vector field term
in Hamiltonian $\frac{e}{c} {\bf A(r)} \cdot {\bf k}$ is replaced
by the circulating superflow velocity as ${\bf v_s (r)} \cdot {\bf
k}$ that causes Doppler shifting of the quasiparticle excitations.
The validity of this method is discussed in length in
Ref.\cite{Mishra} and it is a reliable approximation from the low
to intermediate field regime but could be questionable approaching
$H_{c2}$. Our results, therefore, should be viewed with
reservation at the high field regime where a better method
\cite{BPT} should be applied in principle.

The single-particle Green's function of band $a (=S, L)$ in Nambu
matrix form including Doppler shift of the quasiparticle
excitations is a function of ${\bf r}$ the distance from the
vortex core in addition to the usual momentum ${\bf k}$ and
frequency $\omega$ as follows \cite{Volovik}.

\be
G_{a} ({\bf k, r,\omega})=\frac{[\omega +  {\bf v_s (r)} \cdot
{\bf k}] \tau_0 + \xi_a (k) \tau_3 + \Delta_a \tau_1}{[ \omega +
{\bf v_s (r)} \cdot {\bf k}]^2 - \xi_a ^2 (k) - \Delta_a ^2 }
\ee

\noindent where $\tau_i$ are Pauli matrices and $\xi_a (k)$ is the
quasiparticle energy of band $a$. From the above Green's function
we obtain the local density of states (DOS) of each band as $N_a
(\omega,H,r)= - \frac{1}{\pi} {\rm Tr Im} \sum_k G_{a} ({\bf k,
r,\omega})$. The Doppler shifting energy is given as ${\bf v_s
(r)} \cdot {\bf k}=\frac{k}{m_a} \frac{1}{r} \cos{\theta} = b
\frac{\Delta_L}{\rho} \cos{\theta}$ with normalized distance
$\rho= r/ \xi$ ($\xi=$ coherence length) and "$b$" a constant of
order unity.

We use the above Green's function for $1 \leq \rho \leq R_H /\xi$
with magnetic length $R_H = \alpha \sqrt{\frac{\Phi_0}{\pi H}}$
($\Phi_0$ a flux quanta, $H$ magnetic field, and "$\alpha$"
geometric factor of order unity) because the core region ($\rho <
1$) is thermodynamically negligible if fields is not too close to
$H_{c2}$. Eq.(1) shows that the gaps collapse whenever the Doppler
shifting energy ${\bf v_s (r)} \cdot {\bf k}$ becomes larger than
the gap energy $\Delta_a$. The larger gap $\Delta_L$ does not
collapse even with the largest Doppler shifting of ${\bf v_s}
(\rho=1) \cdot {\bf k_F} \approx \Delta_L$ at the boundary of
vortex core. But the small gap $\Delta_S$ collapses for the region
of $1 < \rho < b \frac{\Delta_L}{\Delta_S}$ to create a finite DOS
$N_S (\omega=0,H,r) \approx N_S ^{normal}$. With this observation,
the magnetic unit cell averaged DOS $\bar{N}_a (\omega,H)=<N_a
(\omega,H,r)>_{cell} =\int_{\xi} ^{R_H} dr^2 N_a (\omega,H,r) /
\pi R_H ^2$ is readily obtained at $\omega=0$ as follows.
\ba \bar{N}_L (\omega=0,H) &=& \frac{0}{\pi R_H ^2} =0 \\
\bar{N}_S (\omega=0,H) &=& N_S ^{normal} \frac{[ (b
\frac{\Delta_L}{\Delta_S})^2 - 1] \xi^2} {R_H ^2} \propto H \ea

Above Eq.(3) holds as far as $\Delta_S < \Delta_L$ and shows that
Volovik effect immediately create a finite DOS with the isotropic
$\pm$s-wave state and there is no threshold value of magnetic
field $H^{*}$ to collapse the small gap $\Delta_S$. Its generic
field dependence is linear in $H$ and its slope is proportional to
$\approx (\frac{\Delta_L}{\Delta_S})^2$. We will show later with
numerical calculations that impurity scattering will smoothen this
generic linear-in-$H$ field dependence and make it more sublinear
and closer to $\propto \sqrt{H}$. Therefore the impurity effect is
important to understand experiments.

The impurity scattering is included by the $\mathcal{T}$-matrix
method, suitably generalized for the $\pm$s-wave pairing model
\cite{Bang-imp}. And impurity induced selfenergies renormalize the
frequencies and OPs as $\omega \rightarrow \tilde{\omega} =\omega
+ \Sigma^{0}_S(\omega) + \Sigma^{0} _L(\omega)$, and $\Delta_{S,L}
\rightarrow \tilde{\Delta}_{S,L} = \Delta_{S,L} + \Sigma^1 _{S}
(\omega) + \Sigma^1 _{L} (\omega),$ with $\Sigma_{S,L} ^{0,1}
(\omega)  = \Gamma \cdot \mathcal{T}^{0,1} _{S,L} (\omega)$ where
$\Gamma= \frac{n_{imp}}{\pi N_{tot}}$; $n_{imp}$ the impurity
concentration and $N_{tot}=N_S ^0 +N_L ^0$ is the total DOS. The
$\mathcal{T}$-matrices $\mathcal{T}^{0,1}$ are the Pauli matrices
$\tau^{0,1}$ components in the Nambu space. We refer readers to
Ref.\cite{Bang-imp} for more details and it is straightforward to
incorporate the local Green's function Eq.(1) with Doppler
shifting into the $\mathcal{T}$-matrix method.

After calculating the averaged $\bar{N}_a (\omega,H)$ for all
frequencies, specific heat is calculated as
\be C_a(T,H) = \int_0 ^{\infty} d \omega \bar{N}_a (\omega, H)
\frac{\omega^2}{T^2} {\rm sech}^2 (\frac{\omega}{2 T}) \ee
Similarly, thermal conductivity is calculated with
\cite{Ambegaoka}
\ba \kappa_a (T,H,r) &\propto & N_a ^0 v_F ^2 \int_0 ^{\infty} d
\omega \frac{\omega^2}{T^2}
{\rm sech}^2 (\frac{\omega}{2 T}) K_a (\omega,T,H,r), \\
K_a (\omega,T,H,r) &=& \frac{1} {Im \sqrt{\tilde{ z}^2 -
\tilde{\Delta}_a ^2}} \times \Big(1 + \frac{|\tilde{ z}|^2 -
|\tilde{\Delta}_a| ^2}{|\tilde{ z}^2 - \tilde{\Delta}_a ^2|} \Big
) \ea
\noindent where $\tilde{z} =\tilde{\omega} +{\bf v}_s ({\bf r)}
\cdot {\bf k_F} $. And then longitudinal and transversal thermal
conductivities are calculated as $\kappa _{\parallel}(T,H) =
\int_{cell} d^2 r \kappa(T,H,r)  / \pi R_H ^2$ and $\kappa^{-1}
_{\perp}(T,H) = \int_{cell} d^2 r \kappa^{-1}(T,H,r)  / \pi R_H ^2
$, respectively.

\begin{figure}
\noindent
\includegraphics[width=90mm]{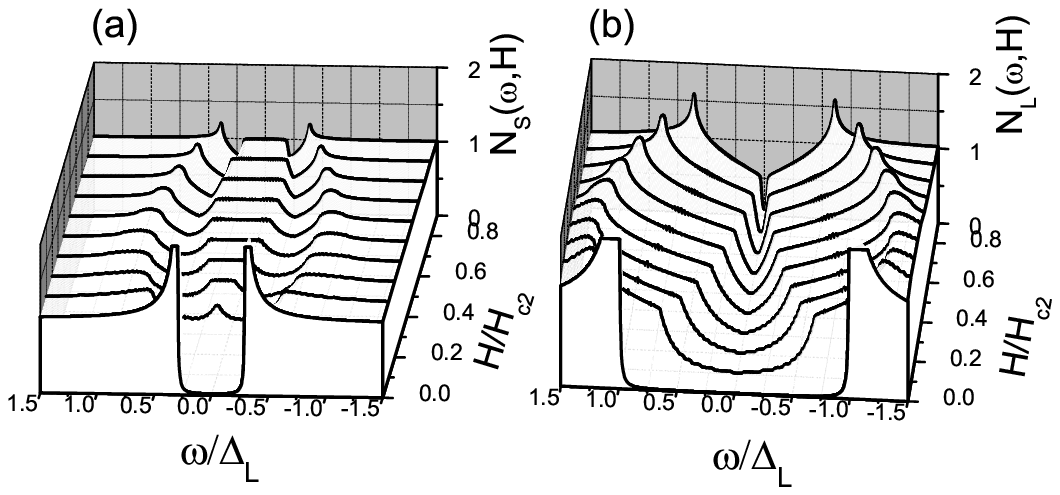}
\includegraphics[width=90mm]{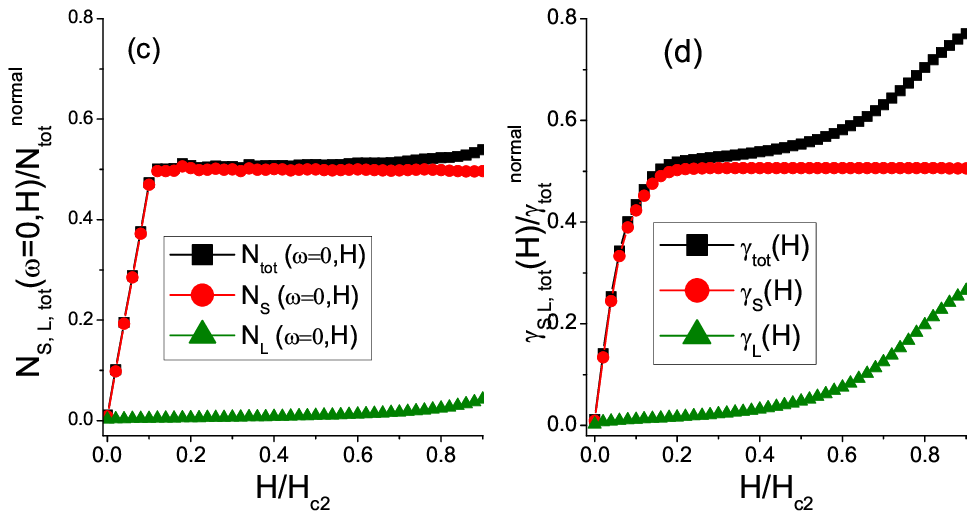}
\caption{(Color online) (a-b) Normalized DOSs with Doppler shifting by
magnetic field $H$, $\bar{N}_{S}(\omega,H)$ and $\bar{N}_{L}(\omega,H)$
for small gap and large gap bands, respectively, in clean limit ($\Gamma/\Delta_L ^0 =0.001$).
(c) Normalized DOSs at $\omega=0$, $\bar{N}_{S, L, tot}(\omega=0,H) /N_{tot} ^{normal}$ of
small gap and large gap bands, and the total, respectively.
(d) Normalized specific heat coefficients $\gamma_{S, L, tot}(H)/ \gamma_{tot} ^{normal}$ of
small gap and large gap bands, and the total contribution, respectively.
\label{fig1}}
\end{figure}

{\it Numerical results and discussions:} Since we consider two
band model, total specific heat and thermal conductivity are the
sum of two contributions from each band. In order to determine the
precise contributions, we need to know the normal state DOSs
$N_{S,L} ^{0}$ as well as Fermi velocity  $v_{F (S,L)}$ of each
band. However, in all numerical calculations in this paper, the
large gap band is almost gapped even with a finite amount of
impurity concentration so that its contribution to specific heat
and thermal conductivity are very small compared to the
contribution from the small gap band. Therefore it is not very
meaningful to determine the precise weighting of each band and we
simply put equal weighting to each band for convenient
illustrations. For the field dependence of the OPs, we use a
phenomenological formula $\Delta_{S,L} (H) = \Delta_{S,L} ^0
\sqrt{1 - H/H_{c2}}$.

\begin{figure}
\noindent
\includegraphics[width=90mm]{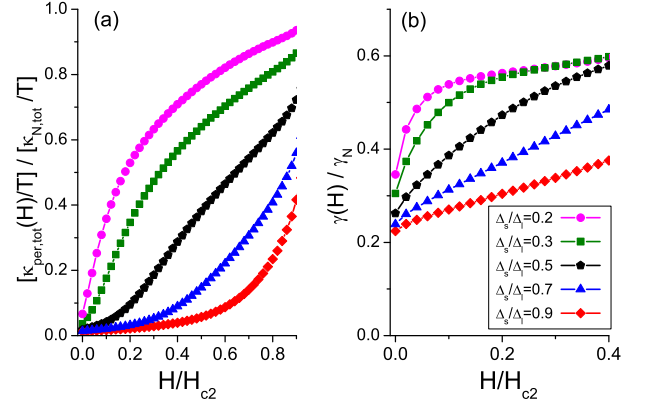}
\caption{(Color online) (a) Normalized total transverse thermal conductivity coefficient
$\lim_{T \rightarrow 0}[\kappa_{\perp,tot}(H)/T] / [\kappa_{N,tot}/T]$ for
different gap size ratios, $|\Delta_S ^0 / \Delta_L ^0|= 0.2, 0.3, 0.5, 0.7$, and $0.9$, respectively.
(b) Normalized specific heat coefficient $\gamma_{tot}(H)/ \gamma_{tot,N}$
for different gap size ratios as in (a). All calculations include the
same concentration of impurities $\Gamma/ \Delta_L=0.05$ with unitary scattering limit ($c=0$)
both for intra- and interband scattering.
\label{fig2}}
\end{figure}

In Fig.1, we show numerical calculations of a pure case
($\Gamma/\Delta_L ^0 =0.001$) with a typical gap size ratio
$|\Delta_S ^0 /\Delta_L ^0|=0.3$. Figures 1(a) and (b) show the
magnetic unit cell averaged DOSs $\bar{N}_{S,L} (\omega, H)$ of
the small and large gap bands, respectively, as a function of
magnetic field $H$. As expected, the small gap DOS
$\bar{N}_S(\omega,H)$ immediately starts collapsing with field and
reaches a constant value $N_S ^{small}$ at $\omega=0$ for $H > 0.1
H_{c2}$ (see Fig.1(c)) while the larger gap DOS
$\bar{N}_{L}(\omega, H)$ remains gapped until $H \rightarrow
H_{c2}$. However, the gapped region in $\bar{N}_{L} (\omega,H)$ in
frequency axis become extremely narrow with increasing field which
will affect specific heat coefficient at high fields (see
Fig.1(d))). Fig.1(c) shows normalized zero energy DOSs
$\bar{N}_{S,L} (\omega=0, H)$ and their total $\bar{N}_{tot} (0,
H)=\bar{N}_{S}(\omega=0, H)+ \bar{N}_{L}(\omega=0, H)$. As
explained above, beyond $H \approx 0.1 H_{c2}$, the small gap
$\bar{N}_{S}(\omega=0, H)$ reaches its full value ($=0.5 N_{tot}
^0$) and the large gap $\bar{N}_{L}(\omega=0, H)$ remains zero
until very close to $H_{c2}$. Fig.1(d) shows normalized specific
heat coefficients $\gamma_{S,L,tot}(H)$, respectively. Some
difference between $\bar{N}_{L}(0, H)$ and $\gamma_{L}(H)$ at the
high field region is due to the finite temperature (we used
$T/\Delta_L ^0 =1/50$ to calculate the $T \rightarrow 0$ limit in
this paper) and very narrowly gapped $\bar{N}_{L} (\omega,H)$ at
the high field region as seen in Fig.1(b).

In Fig.2(a) and (b), we show calculations of total thermal
conductivity and specific heat coefficients with varying gap size
ratio $|\Delta_S ^0 / \Delta_L ^0 | =0.2, 0.3, 0.5, 0.7,$ and 0.9,
respectively. Here we used impurity scattering with $\Gamma /
\Delta_L ^0 =0.05$ and unitary impurity ($c=0$) both for intra-
and interband scattering. These values are not particular choices
for the plots. A wide range of impurity concentrations $\Gamma$
and different strength ($c\neq 0$) of impurity scattering produce
qualitatively similar behaviors.
Fig.2(a) shows normalized total transverse thermal conductivity
$\lim_{T \rightarrow 0} [\kappa_{\perp,S}(H)/T +
\kappa_{\bot,L}(H)/T]/[\kappa_{total}^{normal}/T]$. Only the
transverse $\kappa_{\bot}/T$ (where the external field is applied
as $H \parallel {\bf c}$ and the thermal current is measured in
the plane as ${\bf j}_{th} \| {\bf ab}$) are shown because those
are the measured thermal current in most of experimental settings
\cite{kappa_CoBa122, kappa_KBa122}. The longitudinal thermal
conductivity coefficient $\kappa_{\parallel}/T$ (where ${\bf
j}_{th} \parallel {\bf c}$) behaves a bit more concave, in
particular at the low field regime, because it corresponds to a
parallel circuit of resistors while the transverse one corresponds
to a series circuit. At the higher field regime they become
indistinguishably similar to each other.

The main feature of the results in Fig.2(a) is the systematic
evolution of the slope of $\kappa_{\perp}(H)/T$ at the low field
limit. For a large gap size ratio of $|\Delta_S ^0/\Delta_L
^0|=0.9$, $\lim_{H \rightarrow 0} \kappa_{\perp}(H)/T$ becomes
exponentially flat similar to the behavior of a single gap
isotropic s-wave state. Then with decreasing gap size ratio of
$|\Delta_S ^0 /\Delta_L ^0|$, the low field slope quickly
increases and the overall $H$ dependence of $\kappa_{\perp}(H)/T$
becomes concave down and close to the $\sqrt{H}$ behavior for
$|\Delta_S ^0 /\Delta_L ^0|=0.2$. This concavity can be made even
stronger with decreasing the gap size ratio $|\Delta_S ^0
/\Delta_L ^0|$ and increasing impurity concentration $\Gamma$.

Another important feature is that the values of
$\kappa_{\perp}(H)/T$  (also $\kappa_{\parallel}(H)/T$ although
not shown here) in the zero field limit are negligibly small for
all cases despite substantial impurity scattering; this is even
true with the $|\Delta_S ^0 /\Delta_L ^0|=0.2$ case which shows
the behavior $\approx \sqrt{H}$ as in the d-wave case.
In fact, these extremely small values of the thermal conductivity
coefficient $\kappa_{\perp}(H)/T$ in the zero field limit was
argued as an evidence of an isotropic s-wave gap nature
\cite{kappa_CoBa122}. However, it is a very puzzling feature when
we note that the several experiments \cite{SH_KBa122, SH_CoBa122,
SH_La1111} observed substantial values of the specific heat
coefficients $\gamma(H \rightarrow 0)$ with the same compounds
with similar dopings.
Surprisingly, these seemingly conflicting behaviors are also
obtained in our theoretical calculations. In Fig.2(b), the
normalized specific heat coefficient $\gamma(H \rightarrow 0)/
\gamma_{tot,N}$, with the same parameters as in Fig.2(a), show
substantial values in the zero field limit ($0.5 \gamma_{tot,N}$
is the full value of the small gap band contribution).

It is a common knowledge that the DOS near Fermi level similarly
contributes to both specific heat and thermal conductivity.
However, the discrepancy between $\gamma(H \rightarrow 0)$ and
$\kappa(H \rightarrow 0)/T$ is possible due to the coherence
factor of superconductivity. The kernel of thermal conductivity
(see Eq.(6)), being an energy current-energy current correlation
function, has a destructive coherence factor ("$-$" sign in the
numerator of the last term in Eq.(6)) and the specific heat does
not have such coherence factor. In the case of d-wave pairing, the
same coherence factor becomes very weak in the low energy limit
because the nodal gap $\Delta_D (\theta)$ linearly disappears, so
that $\gamma(H \rightarrow 0)$ and $\kappa(H \rightarrow 0)/T$
behave similarly.

As to the  field dependence of $\gamma(H)$, with a relatively
large gap size ratio, such as $|\Delta_S ^0 / \Delta_L ^0 |=$ 0.7
and 0.9, $\gamma(H)$ is very linear in $H$ for a substantial
region of fields (up to $\approx H_{c2}/2$) despite a finite
$\gamma(H \rightarrow 0)$. Decreasing the gap size ratio to
$|\Delta_S ^0 / \Delta_L ^0 |=$ 0.5, 0.3 and 0.2, the field
dependence of $\gamma(H)$ becomes gradually more concave down.
This behavior is in excellent agreement with the measurements of
Ba$_{0.6}$K$_{0.4}$Fe$_2$As$_2$ \cite{SH_KBa122} ($\approx H$),
(Fe$_{0.92}$Co$_{0.08}$)$_2$As$_2$ \cite{SH_CoBa122} (sublinear in
$H$), and LaO$_{0.9}$F$_{0.1-\delta}$FeAs \cite{SH_La1111}
($\approx \sqrt{H}$).

{\it Conclusion:} Using semiclassical approximation, we have
calculated specific heat coefficient $\gamma(H)$ and thermal
conductivity coefficient $\kappa(H)/T$ of the $\pm$s-wave model in
the mixed state with including impurity scattering. We found that
Doppler shift on quasiparticle excitations on extended states
immediately induce zero energy excitations proportional to $H$
from a smaller gap band. There are no threshold values of small
gap size $\Delta_S ^{*}$ and magnetic field $H^{*}$ to induce such
zero energy excitations. {\it This steep increase of $\gamma(H)$
and $\kappa(H)/T$ linear in $H$ is a generic behavior of any two
gap isotropic s-wave superconductors with unequal sizes of gaps.}
Therefore, our results also provide natural explanation for the
thermal conductivity measurements in MgB$_2$ \cite{MgB2} and
NbSe$_2$\cite{NbSe2}. The sign-changing feature of the $\pm$s-wave
state is not a primary source of the field dependence of
$\gamma(H)$ and $\kappa(H)/T$ but plays an important role through
impurity scattering. The impurity scattering effect with
sign-changing OPs \cite{Bang-imp} efficiently accumulate zero
energy impurity band near Fermi level and reflects it to $\gamma(H
\rightarrow 0)$ (see Fig.2(b)) and makes the field dependence of
$\kappa(H)/T$ and $\gamma(H)$ more concave down and smoothly
curved.

Having found substantial contribution of low energy excitations
from extended state outside of the vortex cores, we justified the
ignoring of the core region contribution {\it a posteriori}. We
compare the numerical calculations to the thermal conductivity
measurement \cite{kappa_CoBa122} of Ba(Fe$_{1-x}$Co$_x$)$_2$As$_2$
and showed that the systematic evolution of the thermal
conductivity coefficient $\kappa(H)/T$ with Co-doping $x$ is well
explained by changing the gap size ratio $|\Delta_S ^0 /\Delta_L
^0|$ in our theoretical calculations, which is also consistent
with the Fermi surface evolution in Ba(Fe$_{1-x}$Co$_x$)$_2$As$_2$
with electron doping by Co in the rigid band picture.
We also showed that the conflicting behaviors between $\gamma(H)$
and $\kappa(H)/T$ in the $H \rightarrow 0$ limit is a natural
consequence of the superconducting coherence factor of thermal
conductivity kernel.
In summary, the $\pm$s-wave state with isotropic s-wave gaps of
unequal sizes can consistently explain the field dependence of
$\gamma(H)$ \cite{SH_CoBa122, SH_KBa122, SH_La1111} and
$\kappa(H)/T$ observed in Co-doped Ba-122 \cite{kappa_CoBa122} and
K-doped Ba-122 \cite{kappa_KBa122}.

{\it Acknowledgement -- } I thank H. H. Wen for useful
discussions. This work was supported by the KOSEF through the
Grants No. KRF-2007-521-C00081.


\begin{references}

\bibitem{Kamihara}
Y.Kamihara et al., J. Am. Chem. Soc., {\bf 130}, 3296 (2008); G.
F. Chen et al., Phys. Rev. Lett. {\bf 100}, 247002 (2008); X. H.
Chen et al., Nature (London) {\bf 453}, 761 (2008).

\bibitem{Mazin}
I.I. Mazin, D.J. Singh, M.D. Johannes, M.H. Du, Phys. Rev. Lett.
{\bf 101}, 057003 (2008); K. Kuroki et al., Phys. Rev. Lett. {\bf
101}, 087004 (2008).

\bibitem{Bang-model}
Y. Bang and H.-Y. Choi, Phys. Rev. B, {\bf 78}, 134523 (2008).

\bibitem{photoemission-s}
H. Ding et al., Euro. Phys. Lett. {\bf 83}, 47001 (2008); T. Kondo
et al., Phys. Rev. Lett. {\bf 101}, 147003 (2008);  L. Wray et
al., Phys. Rev. B 78, 184508 (2008).


\bibitem{1111}
L. Malone et al., Phys. Rev. B 79, 140501(R) (2009);  K. Hashimoto
et al., Phys. Rev. Lett. {\bf 102}, 017002 (2009).

\bibitem{Ba-122}
K. Hashimoto et al., Phys. Rev. Lett. {\bf 102}, 207001 (2009).

\bibitem{T1}
K. Matano et al., Europhys. Lett. {\bf 83}  57001 (2008);  H.-J.
Grafe et al., Phys. Rev. Lett. {\bf 101}, 047003 (2008);  H.
Mukuda et al., J. Phys. Soc. Jpn. {\bf 77} (2008) 093704; Y. Nakai
et al., J. Phys. Soc. Jpn. {\bf 77} (2008) 073701; S. Kawasaki et
al., Phys. Rev. B {\bf 78}, 220506(R) (2008).

\bibitem{BaCo-122}
R. T. Gordon et al., Phys. Rev. B 79, 100506(R) (2009).

\bibitem{theory-T1}
D. Parker, O.V. Dolgov, M.M. Korshunov, A.A. Golubov, I.I. Mazin ,
Phys. Rev. B {\bf 78}, 134524 (2008); A.V. Chubukov, D.V. Efremov,
I. Eremin, Phys. Rev. B {\bf 78}, 134512 (2008); M. M. Parish, J.
Hu, B. A. Bernevig, Phys. Rev. B {\bf 78}, 144514 (2008)

\bibitem{Bang-imp}
Y. Bang, H.-Y. Choi, and H. Won, Phys. Rev. B {\bf 79}, 054529
(2009); Y. Bang, Europhys. Lett. {\bf 86}, 47001 (2009)

\bibitem{SH_KBa122}
G. Mu et al., Phys. Rev. B {\bf 79}, 174501 (2009).

\bibitem{SH_CoBa122}
G. Mu et al., Chin. Phys. Lett. {\bf 27}, 037402 (2010).

\bibitem{SH_La1111}
G. Mu, X. Zhu, L. Fang, L. Shan, C. Ren, and H. Wen, Chin. Phys.
Lett. {\bf 25}, 2221 (2008).

\bibitem{kappa_CoBa122}
M. A. Tanatar et al., Phys. Rev. Lett. {\bf 104}, 067002 (2010);
J. K. Dong et al., arXiv:0908.2209 (unpublished).

\bibitem{kappa_KBa122}
X.G. Luo et al., Phys. Rev. B {\bf 80}, 140503(R) (2009).

\bibitem{Volovik}
G. E. Volovik, JETP Lett. {\bf 58}, 469 (1993); C. Kubert and P.J.
Hirschfeld, Solid State Commun. {\bf 105}, 459 (1998).

\bibitem{Mishra}
V. Mishra, A. Vorontsov, P.J. Hirschfeld, I. Vekhter, Phys. Rev. B
{\bf 80}, 224525 (2009).

\bibitem{BPT}
U. Brandt, W. Pesch, and L. Tewordt, Z. Phys. {\bf 201}, 209
(1967); W. Pesch, Z. Phys. B {\bf 21}, 263 (1975).

\bibitem{Hussey}
N. E. Hussey, Adv. Phys. {\bf 51}, 1685 (2002).

\bibitem{Ambegaoka}
V. Ambegaoka and L. Tewordt, Phys. Rev. {\bf 134}, A805, (1964).

\bibitem{MgB2}
A. V. Sologubenko, J. Jun, S. M. Kazakov, J. Karpinski, H. R. Ott,
Phys. Rev. B, {\bf 66}, 014504 (2002).

\bibitem{NbSe2}
E. Boaknin et al., Phys. Rev. Lett. {\bf 90}, 117003 (2003).


\end{references}
\end{document}